\documentclass{article}
\usepackage{amsfonts,amssymb, amsmath}
\usepackage[english]{babel}

\newtheorem{prop}{Proposition}

\newtheorem{ex}{Example}
\newenvironment{exam}{\begin{ex} \rm }{\end{ex}}

\def\nn{\nonumber }
\def\bq{ \begin{equation}}
\def\eq{ \end{equation}}
\def\ben{ \begin{eqnarray}}
\def\en{ \end{eqnarray}}

\begin{document}


\title{Transformation of the St\"{a}ckel matrices preserving superintegrability}

\author{A.V. Tsiganov \\
\it\small St.Petersburg State University, St.Petersburg, Russia\\
\it\small e--mail: andrey.tsiganov@gmail.com}
\date{}
\maketitle

\begin{abstract}
If we take a superintegrable St\"{a}ckel system and make  variables  "faster" or "slower", that is equivalent to a trivial transformation of the St\"{a}ckel matrix and potentials, then we obtain an infinite family of superintegrable systems with explicitly defined additional integrals of motion. We present some examples of such transformations associated with angle variables expressed via logarithmic functions.
 \end{abstract}

\section{Introduction}
\setcounter{equation}{0}
Let us consider a superintegrable system with closed trajectories. Various time parameterizations of the given trajectories can be associated with different superintegrable systems. The relation between harmonic oscillator and Kepler problem is a main example of such time reparametrizations, which are called Jacobi-Maupertuis transformation, canonical transformation of the extended phase space, St\"{a}ckel  transformation, coupling constant metamorphosis, reciprocal  transformation, etc., see \cite{bk86,gg18,hiet,sb,ts99,ts01}. The main problem here is how to construct additional integrals of motion associated with different times. The purpose of this paper is to discuss trivial transformation of the St\"{a}ckel matrix and potentials which maps the given superintegrable St\"{a}ckel system to an infinite family of superintegrable systems with explicitly defined additional integrals of motion.

A Hamiltonian system on a $2n$-dimensional symplectic manifold $M$ endowed with symplectic form $\Omega$ is said to be completely integrable if it has $n$ smooth functionally independent functions $H_1,\ldots,H_n $ in involution
\[
 \{H_i,H_j \} = 0\,,\qquad i,j=1,\ldots,n
\]
which define Lagrangian submanifold in a phase space \cite{fas05,resh16}.

Integrable system with additional independent integrals of motion $X_1,\ldots,X_k$ commuting with the Hamiltonian $H=H_1$
\bq\label{gen-add-int}
\{H,X_i\}=0\,,\qquad i=1,\ldots,k,
\eq
is the so-called superintegrable system if all the integrals $H_j$ and $X_i$ are well defined functions on  phase space or an open submanifold of phase space.

 Below we suppose that trajectories of superintegrable systems lie on an intersection of $n$-dimensional Lagrangian submanifold with $k$ hyperplanes $X_i=const$. There are also many systems with $n+k$ integrals of motion which have motions linear on tori of smaller dimension $n-k$ \cite{fas05,resh16}. We prefer to study an intersection of algebraic varieties because it allows us to apply standard algebraic geometry methods, for instance Abel's theorem, to construction of integrals of motion.

Properties of the Lagrangian submanifold are described by the Liouville theorem, which implies that almost all points of  phase space are covered by a system of open toroidal domains with the canonical action-angle coordinates
$I_1,\ldots,I_n$ and $\omega_1,\ldots,\omega_n$ so that
\[
\dot{I}_j=0\,,\qquad \dot{\omega}_j=\dfrac{\partial H}{\partial I_j}\,,\qquad
\Omega=\sum_{j=1}^n d I_j\wedge \omega_j\,.
\]
Of course, in general not the entire phase space of a superintegrable system
is fibred by invariant tori of a given dimension $n$ because in fact the orbits form a foliation
with singularities. In order to construct additional integrals of motion it is enough to consider the regular subset of phase space $M$ where any function $X$ independent on $H_1,\ldots,H_n$  has to be some well-defined function on angle variables, see discussion in \cite{op}.

In 1837 Jacobi proved that a product of $n$ smooth plane curves $C_1\times\cdots\times C_n$ is the Lagrangian submanifold in $\mathbb R^{2n}$ \cite{jac37,jac37-2}. Indeed, if we take $n$ plane curves $C_j$ defined by equations of the form
\[
\Phi_j(u_j,p_{u_j},H_1,\ldots,H_n)=0\,,\quad j=1,\ldots,n, \quad \mathrm{det}\left[\dfrac{\partial \Phi_j}{\partial H_k}\right]\neq 0\,,
\]
then solutions of these relations with respect to $H_1,\ldots,H_n$ are in involution according to the Jacobi theorem.

In 1891 St\"{a}ckel found action-angle variables
\bq\label{st-action}
I_i=\sum_{j=1}^n S^{-1}_{ji}\Bigl(p_{u_j}^2+V_j(u_j)\Bigr)\,,\qquad i=1,\ldots,n\,,
\eq
and
\bq\label{st-angle}
\omega_i=\
\dfrac{1}{n}\sum_{j=1}^n\int \dfrac{S_{ij}(u_j)}{\sqrt{\sum_{m=1}^n I_m S_{mj}(u_j)-V_j(u_j)}}\,\mathrm du_j
\eq
for the Jacobi integrable systems with Hamiltonians
\[H=F(I_1,\ldots,I_n)\]
 associated with a product $C_1\times\cdots\times C_n$ of hyperelliptic curves of the form
\[
 C_j:\qquad p_{u_j}^2=\phi_j(u_j,H_1,\ldots,H_n)\equiv\sum_{k=1}^n H_k S_{kj}(u_j)-V_j(u_j)
\]
Here $S$ is the so-called St\"{a}ckel matrix, which $j$ column depends only on coordinate $u_j$, and $V_j(u_j)$ are St\"{a}ckel potentials.
These action-angle variables have been reconstructed in many partial cases by using complete solution of the Hamilton-Jacobi equation and the corresponding generating function or by integration of the form $\theta=\sum p_{u_i}du_i$ along  basic cycles of the corresponding torus, for instance see \cite{gon12,gon14,kal02,op}.

Changing first row in the St\"{a}ckel matrix
\[
S\to\tilde{S}\,,\qquad S_{ij}=\tilde{S}_{ij}\,,\qquad i\neq 1\,,
\]
we make the canonical transformation of the extended phase space
\[H\to\tilde{H}=v^{-1}(H+c)\,,\quad dt\to d\tau=vdt\,,\]
where
\[
v=\frac{\det S}{\det \tilde S}\,,
\]
which always preserves  integrability and, sometimes, superintegrability \cite{bk86,gg18,hiet,sb,ts99,ts01}.

In this note we study transformations which always  preserve superintegrability. Indeed, let us take a superintegrable St\"{a}ckel system with  independent integral of motion
\[
X=F(\omega_1,\omega_2,\ldots,\omega_n)\,.
\]
Here we suppose that action-angle variables exist in a suitable region in phase space,
whereas $X$ is a globally defined function. Transformation of the St\"{a}ckel matrix and potentials
 \[
S\to\tilde{S}=S\cdot\mbox{diag}(k_1^2,k_2^2,\ldots,k_n^2)\,,\qquad\tilde{V}_j= k_j{V}_j\,,\qquad k_i\in\mathbb Z\,,
\]
could preserve  superintegrablity, because the same function on new angle variables
\[
\tilde{X}=F(\tilde{\omega}_1,\tilde{\omega}_2,\ldots,\tilde{\omega}_n)
\]
remains well defined integral of motion. Our main aim is to discuss a few examples of such transformation.

\section{Superintegrable St\"{a}ckel systems}
Trajectories of the St\"{a}ckel system are defined by equations $I_j=\alpha_j$ and $\omega_j=\beta_j$ (\ref{st-action}-\ref{st-angle}) which form the system of Abel's equations
\bq\label{abel-eq}
\dfrac{1}{n}\sum_{j=1}^n\int \dfrac{S_{ij}(u_j)}{\sqrt{\sum_{m=1}^n \alpha_m S_{mj}(u_j)-V_j(u_j)}}\,\mathrm du_j=\beta_i\,,\qquad i=1,\ldots,n\,.
\eq
Solution of this Jacobi inversion problem is simplified if we known algebraic integrals of Abel's equations \cite{bak97}. Various mathematical  constructions of these algebraic integrals were discussed by Euler, Jacobi, Abel, Rishelot, Weierstrass, etc. In mechanics
algebraic integrals of Abel's equations coincide with the globally defined additional integrals of motion $X_i$ (\ref{gen-add-int}), see discussion in \cite{ts09a,ts08,ts09,ts10}.

In 1852 Bertrand proposed  a direct method for searching of dynamical systems with bound trajectories on the plane \cite{bert52} and in 1873 he proved that among central force potentials with bound orbits there are only two types of central force potentials with the property that all bound orbits are also closed orbits \cite{bert73}. In 1877 a similar theorem about central non-potential forces was established by Darboux and Halphen \cite{darb77,hal77}.

 In 1870 Korkin applied Bertrand's method to the investigation of dynamical systems with non-central forces \cite{kork}, which were later generalized by Goryachev, Suslov, Bobylev and Steklov, see textbook \cite{bob} and review \cite{stek}. In contrast with Bertrand's theorem there are infinitely many  systems on the plane with closed trajectories, and now we have a vast literature on superintegrable St\"{a}ckel systems which focuses on construction of potentials producing superintegrable systems with periodic dynamics see, e.g., \cite{gg18,nieto17,kal02,kal10,ts11,mpw13,pw10,ran13,am15}.

In \cite{op} Onofri and Pauri came back to Abel's equations (\ref{abel-eq}) solving them with respect to St\"{a}ckel potentials $V_j$. Instead of  solving Abel's equations we use known solutions which have been collected in  tables of integrals \cite{grr}.

Let us suppose that angle variables
\[
\omega_i=\
\dfrac{1}{n}\sum_{j=1}^n\int \dfrac{S_{ij}(u_j)}{\sqrt{\sum_{m=1}^n I_m S_{mj}(u_j)-V_j(u_j)}}\,\mathrm du_j
\]
 consist only of integrals of algebraic functions. According to Abel's theorem such integrals and, therefore, angle variables have the following form
 \[
 \omega_i=\sum e_j +\sum l_m+\sum \rho_k\,,
 \]
 where $e_j$ are non-elementary by Liouville functions, $l_m$ are logarithmic functions  and $\rho_k$ are rational functions.
 It is quite natural to extract the following four families of superintegrable  St\"{a}ckel systems:
\begin{enumerate}
\item $\omega_i$ is a sum of rational functions $e_j=l_m=0$;
\item $\omega_i$ is a sum of logarithmic and rational functions $e_j=0$;
\item $\omega_i$ is a sum of logarithms $e_j=\rho_k=0$;
\item $\omega_i$ is a sum of non-elementary function $l_m=\rho_k=0$.
\end{enumerate}
Here and below we impose conditions on  each term in the sum $\omega_i$ instead of one condition on function $X=F(\omega_i)$ of this sum. We do not consider possible more general cases .

These four families are related to integration of binomials, trinomials and higher order polynomials
\[
\int R\left(x,\sqrt{a+b x^m}\right)dx,\quad \int R\left(x,\sqrt{ax^2+bx+c}\right) dx,\ldots,\,
\int R\left(x,\sqrt{a_mx^m+\cdots+a_0}\right) dx,
\]
where $R$ is a rational function. All such integrals are well- known, and we can directly take them from the handbook \cite{grr}.

Below we present some examples of the known superintegrable St\"{a}ckel systems which belong to these four families.
\begin{exam} \textbf{ - Sum of rational functions:}
\par\noindent
 Let us start with superintegrable systems separable in Cartesian coordinates $q_ {1,2} $ on the plane. The following
St\"{a}ckel matrix and potentials
\[
S=\left(
 \begin{array}{cc}
 1 & 1 \\
 1 & -1 \\
 \end{array}
 \right)\,,\quad V_1=a q_1^{M_1}\,,\quad V_2=bq_2^{M_2}
\]
give rise to Hamiltonian
\[H=2I_1=p_1^2+p_2^2+ V_1+V_2\]
which commutes with the second angle variable
\bq\label{aa-cart}
\begin{array}{l}
\omega_2=\displaystyle \int \dfrac{dq_1}{\sqrt{I_1+I_2-V_1(q_1)}}-
\displaystyle \int \dfrac{dq_2}{\sqrt{I_1-I_2-V_2(q_2)}}\,,
\end{array}
\eq
involving integrals on differential binomials. In 1853 Chebyshev \cite{cb}  proved that integrals on differential binomials
\[
\int x^m(\alpha+\beta x^n)^pdx\,,
\]
can be evaluated in terms of elementary functions if and only if:
\begin{itemize}
 \item $p$ is an integer,
\item $\dfrac{m+1}{n}$ is an integer,
\item $\dfrac{m+1}{n}+p$ is an integer\,.
\end{itemize}
For the St\"{a}ckel systems
\[\alpha=I_{1,2}\,,\qquad \beta=1\,\qquad m=0, \qquad n=M,\qquad p=-1/2\]
 and, therefore, Hamiltonian
\[
H=p_1^2+p_2^2+ a q_1^{M_1}+bq_2^{M_2}\,,\qquad a,b\in\mathbb R\,,
\]
is superintegrable, if exponents $M_1$ and $M_2$ belong to the following sequence of positive rational numbers
\[
M=0, 1,\dfrac{1}{2},\dfrac{1}{3},\dfrac{1}{4},\cdots, \dfrac{1}{n}\,,\qquad n\in\mathbb{Z}_+\,,
\]
or sequence of negative rational numbers
\[
M=0,-2,-\dfrac{2}{3},-\dfrac{2}{5},-\dfrac{2}{7},\cdots,-\dfrac{2}{2n-1}\,.
\]
The corresponding first integrals $X=(I_1+I_2)^{m_1}(I_1-I_2)^{m_2}\omega_2$ are polynomials with respect to momenta, see details in \cite{ts18,kal02,nieto17}. For instance, at
\[
H=2I_1=p_1^2+p_2^2+ a q_1^{1/5}
\]
second angle variable is the rational function
\[
\omega_2=\frac{5p_1q_1^{4/5}}{a}+\frac{40p_1^3q_1^{3/5}}{3a^2}+
\frac{16p_1^5q_1^{2/5}}{a^3}+\frac{64p_1^7q_1^{1/5}}{7a^4}+\frac{128p_1^9}{63a^5}
+\frac{q_2}{2p_2}\,,
\]
and  additional integral of motion $X=\sqrt{I_1-I_2}\,\omega_2=p_2\omega_2$ is  polynomial in momenta of tenth order.

Discussion of similar dynamical systems with non-potential forces may be found in \cite{bob,kork,stek}.

\end{exam}
\begin{exam} \textbf{- Sum of rational and logarithmic functions:}
\par\noindent
Let us consider superintegrable systems separable in polar coordinates on the plane
\[
r=\sqrt{q_1^2+q_2^2}\,,\qquad \phi=\arctan q_1/q_2.
\]
Standard St\"{a}ckel matrix
\[
S=\left(
 \begin{array}{cc}
 1 & 0 \\
 r^{-2} & -1 \\
 \end{array}
 \right)
\]
gives rise to the action-angle variables
 \bq\label{aa-polar}
 \begin{array}{l}
 I_1=p_r^2+V_1(r)+\dfrac{p_\phi^2+V_2(\phi)}{r^2}\,, \quad
\omega_1=\dfrac12\displaystyle \int\dfrac{rdr}{\sqrt{r^2I_1 -r^2V_1(r)+I_2}}\,, \\
\\
I_2=-p_\phi^2-V_2(\phi)\,,\qquad
\omega_2=-\dfrac12\displaystyle \int\dfrac{dr}{r\sqrt{r^2I_1-r^2V_1(r)+I_2}}+\dfrac12\int\dfrac{d\phi}{\sqrt{I_2-V_2(\phi)}}\,.
\end{array}
\eq
At $V_2(\phi)=0$ and
\bq\label{rpot-log-polar}
 {V}_1(r)=ar^2\qquad\mbox{or}\qquad {V}_1(r)=\dfrac{a}{r}
 \eq
 second angle variable is the sum of logarithmic and rational functions
 \[
 \omega_2=\dfrac{\log A_1+2\mathrm i\phi}{4\sqrt{I_2}}\,,\qquad A_1=\dfrac{2I_2+r^2I_1+2rp_r\sqrt{I_2}}{r^2}
\]
or
\[
\omega_2=\dfrac{\log A_1+\mathrm i\phi}{2\sqrt{I_2}}\,,\qquad A_1= \dfrac{2I_2-ar+2rp_r\sqrt{I_2}}{r}\,.
\]
More generic potentials were considered in \cite{op}.

The following functions on this angle variable are well defined polynomial integrals of motion
\[
X=\exp(4\sqrt{I_2}\,\omega_2)=A_1e^{2\mathrm i \phi}\,,\qquad\mbox{and} \qquad
X=\exp(2\sqrt{I_2}\,\omega_2)=A_1e^{\mathrm i \phi}\,.
\]
Taking real or imaginary part of $X$ one obtains real integrals of motion  which appear in Bertrand's theorem \cite{bert52,bert73}.

For instance, in the second case one gets a Hamiltonian for the superintegrable Kepler system
\[I_1=H=p_r^2+\dfrac{a}{r}+\dfrac{p_\phi^2}{r^2}
\]
with  additional integral of motion
\[
X=\exp(2\sqrt{I_2}\,\omega_2)=\dfrac{(2\mathrm i r p_r\,p_\phi -2p_\phi^2-ar)}{r}\,e^{\mathrm i\phi}\,,
\]
which generates Laplace-Runge-Lenz vector $(Re X,-Im X)$:
\[
Re X=2(p_1q_2-p_2q_1)p_2 -\dfrac{aq_1}{\sqrt{q_1^2+q_2^2}}\,,\quad
-Im X=2(p_1q_2-p_2q_1)p_1+\dfrac{aq_2}{\sqrt{q_1^2+q_2^2}}\,.
\]

If we take one of the potentials from (\ref{rpot-log-polar}) and apply Chebyshev's theorem to the second integral in $\omega_2$ (\ref{aa-polar}), then one gets an action variable of the form
\[
\omega_2=\dfrac{\log A_1+\rho(\phi,p_\phi)}{4\sqrt{I_2}}\,,
\]
where $\rho$ is an algebraic function. The corresponding integral of motion
\[
X=A_1\exp \rho(\phi,p_\phi)
\]
is a transcendental function in momenta. Some examples of superintegrable systems with transcendental integrals of motion may be found in \cite{hiet84}.
\end{exam}

\begin{exam} - \textbf{Sum of logarithmic functions:}
\par\noindent
Let us continue to consider superintegrable systems separable in polar coordinates on the plane. In order to get a desired  sum of the logarithmic function in $\omega_2$ (\ref{aa-polar}) we make standard trigonometric substitution $u=\cos \phi$ in integral
\bq\label{sw-angle}
Z=\dfrac12\int\dfrac{d\phi}{\sqrt{I_2-V_2(\phi)}}=-\dfrac{1}{2}\int\dfrac{du}{\sqrt{(1-u^2)(I_2-V_2(u)}}
\eq
and look over potentials $V_2(u)$ associated with Euler's substitutions
\[\begin{array}{ll}
V_2(u)=\dfrac{\alpha u+\beta}{u^2-1}\,,\qquad &Z=-\dfrac{1}{2}\displaystyle \int \dfrac{du}{\sqrt{-I_2 u^2+\alpha u+(I_2+\beta)}}\\
\\
V_2(u)=\dfrac{\alpha u^2+\beta}{u^2(u^2-1)}\,,\qquad &Z=-\dfrac{1}{2}\displaystyle \int
\dfrac{udu}{\sqrt{-I_2 u^4+(I_2+\alpha)u^2+\beta}}\,.
\end{array}
\]
These integrals are equal to
\[
\begin{array}{l}
Z=\dfrac{1}{2\sqrt{I_2}}\,\log A_2\,,\qquad
 A_2=2I_2\cos\phi+\alpha+2\sqrt{I_2}p_\phi\sin\phi\,,\\
\\
Z=\dfrac{\mathrm i}{4 \sqrt{I_2}}{\,\log A_2}\,,\qquad
A_2= I_2\cos2\phi+\alpha+\sqrt{I_2}p_\phi\sin2\phi\,.
\end{array}
\]
As a result, we obtain angle variable
\bq\label{w-log-ex}
\omega_2=\dfrac{1}{4\sqrt{I_2}}(m_1\log A_1+m_2\log A_2)=\dfrac{1}{\sqrt{I_2}}\log\left( A_1^{m_1}A_2^{m_2}\right)
\eq
and recover the well-known additional integral of motion
\[
X=\exp 4\sqrt{I_2}\omega_2=A_1^{m_1}A_2^{m_2}\,,\qquad m_k=1,2.
\]
for superintegrable system with Hamiltonian
\[
 H=I_1=p_r^2+V_1(r)+\dfrac{p_\phi^2+V_2(\phi)}{r^2}
\]
where $V_1(r)$ is equal to (\ref{rpot-log-polar}) and
\bq\label{vpot-log-polar}
V_2(\phi)=\dfrac{\alpha\cos\phi+\beta}{\cos^2\phi-1}\qquad\mbox{and}\qquad
V_2(\phi)=\dfrac{\alpha\cos^2\phi+\beta}{\cos^2\phi(\cos^2\phi-1)}\,.
\eq
For instance, taking the first potentials from (\ref{rpot-log-polar}) and second from (\ref{vpot-log-polar}) we obtain the
 well-known Smorodinsky-Winternitz superintegrable system \cite{sw} with Hamiltonian
\ben\label{sw0}
H&=&p_1^2+p_2^2+a(q_1^2+q_2^2)+\dfrac{b_1}{q_1^2}+\dfrac{b_2}{q_2^2}\\
\nn\\
&=&p_r^2+\dfrac{p_\phi^2}{r^2}+ar^2+\dfrac{b_1}{r^2\sin^2\phi}+\dfrac{b_2}{r^2\cos^2\phi}\,,\nn
\en
where $b_1=-\alpha-\beta$ and $b_2=-\beta$.

Using another trigonometric substitution $u=\cos k\phi$ one gets superintegrable systems with additional integrals of motion which are higher order polynomials in momenta. We discuss these systems in the next Section.
\end{exam}

\begin{exam} - \textbf{Sum of elliptic integrals}:
\par\noindent
Let us consider superintegrable system separable in elliptic coordinates on the plane $u_{1,2}$ defined by
\[
1-\dfrac{q_1^2}{\lambda-\kappa}-\dfrac{q_2^2}{\lambda+\kappa}=\dfrac{(\lambda-u_1)(\lambda-u_2)}{\lambda^2-\kappa^2}\,,\qquad \kappa\in \mathbb R\,.
\]
The corresponding St\"{a}ckel matrix
\bq\label{ell-srel}
S=\left(
 \begin{array}{cc}
 \dfrac{u_1}{u_1^2-\kappa^2} & \dfrac{u_2}{u_2^2-\kappa^2} \\
 \\
 \dfrac{1}{u_1^2-\kappa^2} & \dfrac{1}{u_2^2-\kappa^2} \\
 \end{array}
 \right)\,,\qquad \
\eq
\end{exam}
determines action variable
\[
I_1=\dfrac{(u_1^2-\kappa^2)(p_{u_1}^2+V_1)}{u_1-u_2}+\dfrac{(u_2^2-\kappa^2)(p_{u_2}^2+V_2)}{u_2-u_1}
\]
and angle variable
\bq\label{w-ell-ex}
\omega_2=e_1+e_2\,,\qquad e_i=\dfrac{1}{2}\int \dfrac{du_i}{\sqrt{(u_i^2-\kappa^2)\Bigl(u_i I_1+I_2-V_i(u_i)(u_i^2-\kappa^2)\Bigr)}}
\,,
\eq
which is a sum of elliptic integrals at
\[
V_j(u)=4\left(a+\dfrac{b_1}{(\kappa-u)^2}+\dfrac{b_2}{(\kappa+u)^2}\right)
\]
 Using standard divisor arithmetics on the corresponding elliptic curve, we can easily obtain an additional integral of motion for the
 Smorodinsky-Winternitz superintegrable system \cite{sw}
\ben\label{sw0-ell}
H&=&p_1^2+p_2^2+a(q_1^2+q_2^2)+\dfrac{b_1}{q_1^2}+\dfrac{b_2}{q_2^2}\\
\nn\\
&=&\dfrac{(u_1^2-\kappa^2)p_{u_1}^2-(u_2-\kappa^2)p_{u_2}^2}{4(u_1-u_2)}+a(u_1+u_2)
-\frac{2\kappa b_1}{(\kappa-u_1)(\kappa-u_2)}+\frac{2\kappa b_2}{(\kappa+u_1)(\kappa+u_2)}
\,,\nn
\en
which coincides with Euler algebraic integral for Abel's equations, see details in \cite{ts09,ts12}.

In the next Section we discuss transformation of the St\"{a}ckel matrix which allows us to get an infinite family of superintegrable systems starting with the known superintegrable St\"{a}ckel system.

\section{Transformation of the St\"{a}ckel matrix}
Our main idea is to take a superintegrable St\"{a}ckel system and make some variables "faster" or "slower". Indeed, multiplying columns of the St\"{a}ckel matrix and potentials on integers
\[ \tilde{S_{ij}}={k^2_j}{S}_{ij}\,,\qquad\tilde{V}_j= k_j{V}_j\,,\qquad k_j\in \mathbb Z\,,\]
we change separated relations
\[
\begin{array}{rcl}
p_{u_j}^2&=& \sum I_i S_{ij}(u_j)-V_j(u_j)\,,\\
\\
k_j^{-2}p_{u_j}^2&=&\sum I_i S_{ij}(u_j)- V_j(u_j)\,,
\end{array}
\]
and angle variables
\[
\begin{array}{rcl}
\omega_j&=&\displaystyle
\dfrac{1}{n}\left(\int\dfrac{du_1}{p_{u_1}}+\int\dfrac{du_2}{p_{u_2}}+\cdots+\int\dfrac{du_n}{p_{u_n}}\right)
\\ \\
\tilde{\omega}_j&=&\displaystyle
\dfrac{1}{n}\left({ k_1}\int\dfrac{du_1}{p_{u_1}}+{ k_2}\int\dfrac{du_2}{p_{u_2}}+\cdots
+{ k_n}\int\dfrac{du_n}{p_{u_n}}\right)\,.
\end{array}
\]
In fact, this transformation of the St\"{a}ckel matrix and potentials is equivalent to non-canonical transformation of momenta $p_{u_j}\to k_j^{-1}p_{u_j}$\,.

If angle variable $\omega_j$ is a sum of  rational function, then variable $\tilde{\omega}_j$ is also a sum of  rational function
\[
\begin{array}{l}
\omega_j=\rho_1+\rho_2+\cdots+\rho_n=R_j
\\ \\
\tilde{\omega}_j=k_1\rho_1+k_2\rho_2+\cdots+k_n\rho_n=\tilde{R}_j\,,
\end{array}
\]
which can be also used for construction of additional integrals of motion which are polynomials in momenta.

If $\omega_j$ is a sum of logarithmic functions, then variable $\tilde{\omega}_j$ is also a sum of logarithms
\bq\label{k-add-log}
\begin{array}{l}
\omega_j=g(I)\left(\ln A_1+\ln A_2+\cdots+\ln A_n\right)=g(I)\ln \left( A_1A_2\cdots A_n\right)
\\ \\
\tilde{\omega}_j=g(I)\left(k_1\ln A_1+k_2\ln A_2+\cdots+k_n\ln A_n\right)=g(I)\ln \left( A_1^{k_1}A_2^{k_2}\cdots A^{k_n}_n\right)\,,
\end{array}
\eq
where $g(I)$ is a function on action variables as in (\ref{w-log-ex}). In this case transformation of the St\"{a}ckel matrix
changes the original additional integral of motion
\[
X=A_1A_2\cdots A_n \to \tilde{X}=A_1^{k_1}A_2^{k_2}\cdots A^{k_n}_n\,,
\]
that allows us to get superintegrable systems with additional integrals of motion which are polynomials of higher order in momenta.

If $\omega_j$ is a sum of elliptic (hyperelliptic) integrals as in (\ref{w-ell-ex}), then divisors arithmetics can also be applied
to $\tilde{\omega}_j$
\[
\omega_j=e_1+\cdots+ e_n\,,\qquad \tilde{\omega}_j=k_1e_1+\cdots+ k_ne_n\,,
\]
see discussion of addition and multiplication of elliptic integrals in \cite{hand06}. In this case  well-defined additional integrals of motion $X$ could be rational functions in momenta. We plan to discuss the relations of superintegrable systems and arithmetic of divisors in further publications.

Thus, we suppose that the following Proposition can be true.
\begin{prop} Transformation of the St\"{a}ckel matrix and potentials
 \bq\label{st-trans}
S\to\tilde{S}=S\cdot\mbox{diag}(k_1^2,k_2^2,\ldots,k_n^2),\qquad
V_j\to \tilde{V}_j= k_j{V}_j\,,\qquad k_j\in\mathbb Z
\eq
preserves superintegrability.
 \end{prop}
We can prove this Proposition only  in partial cases at $n=2,3$.

Transformation (\ref{st-trans}) changes diagonal metric $g_{jj}\to\tilde{g}_{jj}=g_{jj}k_i^{-2}$ in the original Hamiltonian
\[
H\to \tilde{H}=\tilde{I}_1=\sum \tilde{\mathrm g}_{jj}\,p_{u_j}^2+{V}(u_1,\ldots,u_n)=\sum \mathrm g_{jj}\left(\dfrac{p_{u_j}}{k_j}\right)^2\,+V(u_1,\ldots,u_n)\,.
\]
After this transformation the new metric could be
\begin{itemize}
 \item equivalent to the original metric;
 \item non-equivalent to the original metric.
\end{itemize}
In \cite{kal10} one  can find examples of such transformations in $\mathbb R^3$, which transform flat metric to the non-flat metric and vise versa. Here we consider only some examples of superintegrable St\"{a}ckel systems with two degrees of freedom.

\begin{exam} - \textbf{System separable in Cartesian coordinates}
\par\noindent
In this case transformation of the St\"{a}ckel matrix (\ref{st-trans})
 \[
S=\left(
 \begin{array}{cc}
 1 & 1 \\
 1 & -1 \\
 \end{array}
 \right)\to
 \tilde{S}=\left(\begin{array}{cc}
 k_1^2& k_2^2\\
 k_1^2& -k_2^2
 \end{array}\right)
\]
 does not change properties of the original metric
 \[
\mathrm g=\left(
 \begin{array}{cc}
 1 & 0 \\
 0 & 1 \\
 \end{array}
 \right)\to \tilde{\mathrm g}=\left(
 \begin{array}{cc}
 k_1^{-2} & 0 \\
 0 & k_2^{-2} \\
 \end{array}
 \right)
 \]
 due to the existence of additional canonical transformation
 $p_i\to k_ip_i$ and $q_i\to k_i^{-1}q_i$, which allows us to change the potential instead of metric.

For instance, Hamiltonian of the Smorodinsky-Winternitz superintegrable system
\[
H=p_1^2+p_2^2+a(q_1^2+q_2^2)+\frac{b_1}{q_1^2}+\frac{b_2}{q_2^2}
\]
 after transformation (\ref{st-trans}) looks like
\[
\tilde{H}=\left(\frac{p_1}{k_1}\right)^2+\left(\frac{p_2}{k_2}\right)^2+a(q_1^2+q_2^2)+\frac{b_1}{q_1^2}+\frac{b_2}{q_2^2}\,.
\]
After canonical transformation $p_i\to k_ip_i$ and $q_i\to k_i^{-1}q_i$ one gets Hamiltonian
\bq\label{sw-cart}
\tilde{H}=p_1^2+p_2^2+ak_1^{-2}q_1^2+ak_2^{-2}q_2^2+\frac{k_1^2b_1}{q_1^2}+\frac{k_2^2b_2}{q_2^2}
\eq
of the well-known superintegrable system with integral of motion which is higher order polynomial in momenta associated with logarithmic angle variable $\tilde{\omega}_2$ of the form (\ref{k-add-log}) \cite{ts18,mpw13}.
\end{exam}

\begin{exam} - \textbf{System separable in polar coordinates}
\par\noindent
In this case transformation of the St\"{a}ckel matrix (\ref{st-trans})
\[
S=\left(
 \begin{array}{cc}
 1 & 0 \\
 r^{-2} & -1 \\
 \end{array}
 \right)
\to \tilde{S}
=\left(
 \begin{array}{cc}
 k_1^2 & 0 \\
 k_1^2r^{-2} & -k_2^2 \\
 \end{array}
 \right)
\]
also does not change properties of the original metric
\[
\mathrm g=\left(
 \begin{array}{cc}
 1 & 0 \\
 0 & r^{-2} \\
 \end{array}
 \right)\to
\tilde{ \mathrm g}=\left(
 \begin{array}{cc}
 k_1^{-2} & 0 \\
 0 & k_2^{-2}r^{-2} \\
 \end{array} \right)\,.
\]
For instance, let us consider superintegrable Hamiltonians
\[
 H=I_1=p_r^2+\dfrac{p_\phi^2}{r^2}+V_1(r)+\dfrac{V_2(\phi)}{r^2}
\]
where St\"{a}ckel potentials $V_{1,2}$ are given by (\ref{rpot-log-polar}) and (\ref{vpot-log-polar}). After transformation (\ref{st-trans}) we obtain the following Hamiltonian
\[
\tilde{H}=\left(\frac{p_r}{k_1}\right)^2+\frac{1}{r^2}\left(\frac{p_\phi}{k_2}\right)^2+V_1(r)+\dfrac{V_2(\phi)}{r^2}\,.
\]
Canonical transformation $r\to k_1^{-1}r$ and $\phi\to k_2^{-1}\phi$ reduces metric to the original one and changes potential
\[
\tilde{H}=p_r^2+\frac{p_\phi^2}{r^2}+V_1(k_1^{-1}r)+\dfrac{V_2(k_2^{-1}\phi)}{k_1^{-2}r^2}\,.
\]
As a result, one gets the well-known superintegrable systems with additional integrals of motion $X$, which are polynomials in momenta of higher order associated with logarithmic angle variable $\tilde{\omega}_2$ of the form (\ref{k-add-log}), see \cite{gon12, kal10,pw10,ran13,ttw10}.

For instance, we can reconstruct a second family of superintegrable Hamiltonians associated with Smorodinsky-Winternitz superintegrable system
(\ref{sw0})
\bq\label{sw-pol}
\tilde{H}=p_r^2+\dfrac{p_\phi^2}{r^2}+ar^2+\dfrac{b_1}{r^2\sin^2 k\phi}+\dfrac{b_2}{r^2\cos^2 k\phi}\,,
\eq
 which can be also obtained using trigonometric substitution $u=\cos k\phi$ in (\ref{sw-angle}).
\end{exam}

\begin{exam} - \textbf{Systems separable in parabolic coordinates}
\par\noindent
Let us consider superintegrable systems separable in parabolic coordinates $u_{1,2}$ on the plane defined by
\[
q_1=u_1u_2,\qquad q_2 =(u_1^2-u_2^2)/2\,.
\]
The corresponding St\"{a}ckel matrix
\[
S=\left(
 \begin{array}{cc}
 u_1^2 & u_2^2 \\
 1 & -1 \\
 \end{array}
 \right)\,,
\]
determines action variables
\[
\begin{array}{l}
I_1=\frac{p_{u_1}^2+V_1(u_1)+p_{u_2}^2+V_2(u_2)}{u_1^2+u_2^2}\,,\quad
I_2=\frac{u_2^2\bigl(p_{u_1}^2+V_1(u_1)\bigr)-u_1^2\bigl(p_{u_2}^2+V_2(u_2)\bigr)}{u_1^2+u_2^2}
\end{array}
\]
and angle variables
\[
\begin{array}{l}
\omega_1=-\frac12\left(
{\displaystyle\int} \frac{u_1^2du_1}{\sqrt{u_1^2I_1-V_1(u_1)+I_2}}
+{\displaystyle\int} \frac{u_2^2du_2}{\sqrt{u_2^2I_1-V_2(u_2)-I_2}}
\right)\\ \\
\omega_2=-\frac12\left(
{\displaystyle\int}\frac{du_1}{\sqrt{u_1^2I_1-V_1(u_1)+I_2}}
-{\displaystyle\int} \frac{du_2}{\sqrt{u_2^2I_1-V_2(u_1)-I_2}}
\right)\,.
\end{array}
\]
In order to get a desired sum of logarithms we reduce integrals to the following form
\[
\displaystyle \int R\left(x,\sqrt{ax^2+bx+c}\right) dx\,,
\]
where $R$ is a rational function, and apply the so-called Euler's substitutions. One gets logarithms for the following potentials only:
\[
V(u)=au^2+bu+c\qquad \Rightarrow\qquad \displaystyle{\int} \dfrac{du}{\sqrt{(I_1-a)u^2-bu+I_2-c}}\]
and
\[
V'(u)=\dfrac{au^4+bu^2+c}{u^2}\qquad \Rightarrow\qquad \displaystyle{\int} \dfrac{udu}{\sqrt{(I_1-a)u^4+(I_2-b)u^2-c}}\,.
\]
The corresponding Hamiltonian commutes with second angle variable
\[
\omega_2=\frac{1}{2\sqrt{I_1-a_1}}\log A_1+\frac{1}{2\sqrt{I_1-a_2}}\log A_2\,,
\]
which is the sum of two logarithms with a common factor at $a_1=a_2=const$ only. This constant is irrelevant since it can always be
associated with a shift of $I_1$, so we put $a_1=a_2=0$. As a result we have three superintegrable
Hamiltonains
\bq\label{ham-par}
H^{(j)}=I_1=\frac{p_{u_1}^2+p_{u_2}^2}{u_1^2+u_2^2}+ U_j(u_1,u_2)\,,\qquad j=1,2,3
\eq
where
\[
\begin{array}{c}
U_1=\frac{V(u_1)+V(u_2)}{u_1^2+u_2^2}\,,\qquad U_2=\frac{V(u_1)+V'(u_2)}{u_1^2+u_2^2}\,,\qquad
U_3=\frac{V'(u_1)+V'(u_2)}{u_1^2+u_2^2}\,.
\end{array}
\]
 Additional integrals of motion are equal to
\[
X_1=\exp(2\sqrt{I_1}\omega_2)= A_1A_2\,,\quad X_2=A_1A'_2\,,\quad X_3= A'_1A'_2\,,
\]
where
\[
\begin{array}{c}
A_j=\frac{2I_1u_j+2\mathrm i p_{u_j}\sqrt{I_1}-b_j}{2\sqrt{I_1}}\quad\mbox{and}\quad
 A'_j=\left(\frac{2 I_1 u_j^2+I_2+2\mathrm i u_jp_{u_j}\sqrt{I_1}-b_j}{2\sqrt{I_1} }\right)^2\,.
 \end{array}
\]
Taking real or imaginary part of $X_k$ one obtains real integrals of motion functionally independent on action variables $I_{1,2}$. Obviously, in this way we produce only one new independent integral for each Hamiltonian.

Transformation of the St\"{a}ckel matrix
\[
S=\left(
 \begin{array}{cc}
 u_1^2 & u_2^2 \\
 1 & -1 \\
 \end{array}
 \right)\to
 \tilde{S}=\left(\begin{array}{cc}
 k_1^2u_1^2& k_2^2u_2^2\\
 k_1^2& -k_2^2
 \end{array}\right)
\]
changes metric in Hamiltonians (\ref{ham-par})
\[
\mbox{g}=\frac{1}{u_1^2+u_2^2}\left(
 \begin{array}{cc}
 1 & 0 \\
 0 & 1
 \end{array}
 \right)\to\tilde{\mbox{g}}=\frac{1}{u_1^2+u_2^2}\left(
 \begin{array}{cc}
 k_1^{-2} & 0 \\
 0 & k_2^{-2}
 \end{array}
 \right)
\]
second angle variable (\ref{k-add-log}) and
additional integrals of motion
\[
\tilde{X}_1= A_1^{k_1}A_2^{k_2}\,,\quad \tilde{X}_2=A_1^{k_1}{A'_2}^{k_2}\quad \tilde{X}_3={ A'_1}^{k_1}{A'_2}^{k_2}\,.
\]
Taking real or imaginary part of $\tilde{X}_k$ we obtain additional integrals of motion which are polynomials in momenta of order $m=3,4,\ldots$. Such superintegrable systems have not been researched in the literature before.
\end{exam}

\begin{exam} - \textbf{Systems separable in elliptic coordinates}
\par\noindent
Let us continue to study Hamilton-Jacobi equation $H=E$ for the Smorodinsky-Winternitz superintegrable system which admits separation of variables in Cartesian, polar and elliptic coordinates on the plane.

In elliptic coordinates Hamiltonian (\ref{sw0-ell}) is equal to
\[
H=\frac{(u_1^2-\kappa^2)p_{u_1}^2-(u_2-\kappa^2)p_{u_2}^2}{4(u_1-u_2)}+a(u_1+u_2)
-\frac{2\kappa b_1}{(\kappa-u_1)(\kappa-u_2)}+\frac{2\kappa b_2}{(\kappa+u_1)(\kappa+u_2)}\,.
\]
Using transformation (\ref{st-trans}) one gets new Hamiltonians
\ben\label{sw-ell}
\tilde{H}&=&\frac{u_1^2-\kappa^2}{4(u_1-u_2)}\left(\dfrac{p_{u_1}}{k_1}\right)^2
-\frac{u_2-\kappa^2}{4(u_1-u_2)}\left(\dfrac{p_{u_2}}{k_2}\right)^2
\\
\nn\\
&+&a(u_1+u_2)
-\frac{2\kappa b_1}{(\kappa-u_1)(\kappa-u_2)}+\frac{2\kappa b_2}{(\kappa+u_1)(\kappa+u_2)}\,.\nn
\en
commuting with a second action variable $I_2$. At $k_1=1,2,3$ and $k_2$ we can  also get additional integral of motion $X$ associated with angle variable $\omega_2$ (\ref{w-ell-ex}) using explicit formulae for doubling and tripling  divisors on elliptic curves from \cite{hand06}. In contrast with (\ref{sw-cart}) and (\ref{sw-pol}) additional integral $X$ is the globally defined rational function on the projective plane.

Summing up, we suppose that the well-known Smorodinsky-Winternitz superintegrable system belongs to the three infinite families of superintegrable systems (\ref{sw-cart}), (\ref{sw-pol}) and (\ref{sw-ell}).

\end{exam}

\begin{exam} - \textbf{System with masses depending on coordinates}
\par\noindent
Let us consider superintegrable Hamiltonian
\[
H=I_1=(q_1^2+q_2^2)^3(p_1^2+p_2^2)+aq_1
\]
with masses depending on coordinates \cite{am15}. Substituting this Hamiltonian into the computer program \cite{ts05} one gets variables of separation
\[
u_1=\dfrac{q_1^2}{(q_1^2+q_2^2)^2}\,,\qquad u_2= -\frac{q_2^2}{(q_1^2+q_2^2)^2}\,,
\]
characteristic Killing tensor generating second quadratic integral of motion, St\"{a}ckel matrix
\[
S=\left(
 \begin{array}{cc}
 \frac14 & \frac14 \\ \\
 \frac1{u_1} & \frac1{u_2}
 \end{array}
\right)
\]
second angle variable
\[
\omega_2=-\frac{1}{\sqrt{I_1}}\Bigl(2 \log A_1 +\log A_2 \Bigr)
\]
commuting with $H=I_1$ and the globally defined additional integral of motion
\[
X=\exp\left(-\sqrt{I_1}\omega_2\right)=A_1^2A_2\,,
\]
where
\[
\begin{array}{c}
A_1=\frac{ I_1\sqrt{u_1}+2\mathrm i p_{u_1}\sqrt{u_1}\sqrt{I_1}-a/2 }{ \sqrt{I_1}}\,,\quad
A_2=\frac{I_1u_2+2I_2+2\mathrm i p_{u_2}u_2\sqrt{I_1}}{ \sqrt{I_1}}\,.
\end{array}
\]
After transformation of the St\"{a}ckel matrix (\ref{st-trans})
\[
S=\left(
 \begin{array}{cc}
 \frac14 & \frac14 \\ \\
 \frac1{u_1} & \frac1{u_2}
 \end{array}
\right)\to \tilde{S}=\left(
 \begin{array}{cc}
 \frac{k_1^2}{4} & \frac{k_2^2}{4} \\ \\
 \frac{k_1^2}{u_1} & \frac{k_2^2}{u_2}
 \end{array}
\right)
\]
one gets superintegrable Hamiltonian
\[
\tilde{H}=\tilde{I}_1=\frac{4u_1}{u_1-u_2}\left(\frac{p_{u_1}}{k_1}\right)^2+\frac{4u_2}{u_2-u_1}\left(\frac{p_{u_2}}{k_2}\right)^2
+\frac{a\sqrt{u_1}}{u_1-u_2}
\]
with additional integral of motion
\[
\tilde{X}=\exp\left(-\sqrt{\tilde{I}_1}\tilde{\omega}_2\right)=A_1^{2k_1}A_2^{k_2}\,.
\]
which is independent from action variables. Taking real or imaginary part of $\tilde{I}^{2k_1+k_2}\tilde{X}$ we obtain two real integrals of motion, which are polynomials in momenta.

In similar manner we can get infinite families of superintegrable systems associated with superintegrable systems with masses depending on coordinates from \cite{gg18,ran16,am15}
\end{exam}

\section{Conclusion}
 For the given St\"{a}ckel matrix $S$ we can calculate all the potentials $V_j$ for which angle variables (\ref{st-action}) consist of rational, logarithmic and non-elementary functions. Then we can apply transformation of the St\"{a}ckel matrix and potentials (\ref{st-trans}) to construct infinite families of superintegrable systems. For instance, we can start with St\"{a}ckel matrices $S$ associated with orthogonal curvilinear coordinate systems in spaces of constant curvature from \cite{olev}. For instance,  there are 4 coordinate systems and St\"{a}ckel matrices for $\mathbb E^2$ ;\quad 2 systems for $\mathbb S^2$;\quad 9 systems for $\mathbb H^2$ and 11 systems for $\mathbb E^3$;\quad 6 systems for $\mathbb S^3$;\quad 45 system in $\mathbb H^3$. In this paper we present some examples of such construction including Smorodinsky-Winternitz system, which generates three infinite families of superintegrable systems.

 More complicated construction of superintegrable systems involves non-point canonical transformation relating original variables and variables of separation. If matrix $S$ and potentials $V_j$ depend on parameter $a$, then action-angle variables (\ref{st-action}-\ref{st-angle}) also depend on this parameter. After non-point canonical transformation
\[
q_i=f_i(u,p_u,a)\,,\qquad p_i=g_i(u,p_u,a)
\]
angle variables $\omega_i(q,p,a)$ could be used to construct an additional integral of motion $X=F(\omega_i)$ which becomes a
globally defined function only at $a\to 0$. Examples of such superintegrable systems may be found in \cite{cr13,pw11,ts17}. It will be interesting to apply transformation of the St\"{a}ckel matrix and potentials (\ref{st-trans}) to these systems.

The work was supported by the RFBR grant (project 18-01-00916).

\end{document}